\documentclass[lettersize,journal]{IEEEtran}
\usepackage{amsmath,amsfonts}
\usepackage{algorithmic}
\usepackage{algorithm}
\usepackage{array}
\usepackage[caption=false,font=normalsize,labelfont=sf,textfont=sf]{subfig}
\usepackage{textcomp}
\usepackage{stfloats}
\usepackage{url}
\usepackage{verbatim}
\usepackage{graphicx}
\usepackage{cite}
\begin{document}

\title{Pitch Estimation with Mean Averaging Smoothed Product Spectrum and Consonance Evaluation Using MASP}

\author{M. Y. Baskin~
\thanks{}
\thanks{}}

\markboth{1}%
{Shell \MakeLowercase{\textit{et al.}}: A Sample Article Using IEEEtran.cls for IEEE Journals}


\maketitle

\begin{abstract}
This study introduces Mean Averaging Smoothed Product (MASP) Spectrum, which is a modified version of the Harmonic Product Spectrum, designed to enhance pitch estimation for many algorithm-wise deceptive frequency spectra that still lead clear pitches, for both harmonic and inharmonic cases. By introducing a global mean based smoothing for spectrum, the MASP algorithm diminishes the unwanted sensitivity of HPS for spectra with missing partials. The method exhibited robust pitch estimations consistent with perceptual expectations. Motivated upon the strong correlation between consonance and periodicity, the same algorithm is extended and, with the proposition of a harmonicity measure (H), used to evaluate musical consonance for two and three tones; yielding consonance hierarchies that align with perception and practice of music theory. These findings suggest that perception of pitch and consonance may share a similar underlying mechanism that depends on spectrum.
\end{abstract}


\section{Introduction}
\IEEEPARstart{P}{itch} perception is one of the central subjects of psychoacoustics that aims to precisely define what the perceived pitch of a complex tone is, and what procedure a complex waveform is undergone to attain the perceived pitch as output. The modern methods primarily fall under the umbrellas of two main categories: Autocorrelation and Pattern-Matching [1] (also referred as \textit{temporal} and \textit{spectral} models). Those two categories fundamentally differ regarding how they treat a complex tone. Autocorrelation methods directly use the time series waveform to evaluate the pitch, by constructing a similarity/difference measure between a frame of the waveform and its time-shifted self. Pattern-matching models, however, use frequency spectrum of the waveform evaluated for a frame of the waveform. Although autocorrelation models are much more computing-efficient due to both simplicity of the model and a direct link to Fast Fourier Transform, pattern-matching models are still noteworthy in psychoacoustics regard, since, which one (or if both) is the true underlying mechanism behind the auditory system is still a matter of discussion.

\section{Background}
\subsection{Definition of the Problem, Problem of Definition}
Plack and Oxenham [2] define pitch as “the quality that makes it possible to judge sounds as ‘higher’ or ‘lower’ in the sense associated with musical melodies.”

Including the definition above, the verbal definitions lay a clear foundation that there is no intuitive disagreement of what pitch is. However, it is quite the contrary as far as a rigorous mathematical definition is concerned.

\subsubsection{Intuitive definition vs scientific definition}
“...pitch is an attribute of sensation.” says Plack and Oxenham [2]. The core problem with the scientific definition is that the only robust evidence of the existence of the concept of pitch is testimony, rather than the intrinsic features of \textit{cochlea}/\textit{cortex}/\textit{brain}, that we know for sure by observation, that shows how pitch is constructed given a waveform. Hence all the attempts to construct a pitch perception algorithm are nothing but pattern recognition models that are attempted to fit '\textit{pitch - frequency spectrum}' (in pattern-matching) or '\textit{pitch - waveform}' (in autocorrealation) pairs based on hearing experiments.

\subsubsection{Frequency vs pitch}
In the contexts outside of psychoacoustics, including music theory, the term frequency is loosely \& interchangeably used with pitch. However this is never the case regarding pitch perception theories literature. A complex waveform, regardless harmonic or inharmonic, is composed of sinusoids. Frequency, or frequency spectrum, refers to the inverse periods of each individual sinusoidal wave. Pitch is, however, what we perceive as ‘high’ or ‘low’ of the overall waveform. They, of course, are the same for pure sinusoids.

\subsubsection{Ambiguities in pitch perception}
There are cases where perceptually not a single clear pitch is heard [3], for mostly inharmonic tones. Most of the pitch perception algorithms however end up either 1) assigning a single number as a pitch as a results of the frequency spectrum undergoing a model, or 2) creating a mathematically ambiguous ‘pitch spectrum’ that require further human interpretation for evaluation of the pitch, which spoils the entire point of building an algorithm.

A noteworthy example of pitch ambiguity, which can even be considered harmonic, is \textit{Shepard Tone} [4]. A tone could be heard as infinitely rising which is impossible to explain with a pitch model discretely defined on a linear scale.

\subsection{Common Pitch perception Algorithms and Common Errors}

\subsubsection{Lowest partial model}
A sawtooth wave of 100 Hz is an ideal example of a harmonic complex waveform, that yields a clear perception of pitch. It is composed of  sinusoids of frequencies that are integer multiples of 100 Hz, ideally up to infinity, whose amplitudes decay proportionally to harmonic number.  Inspired by that, one would assume that the lowest-in-frequency partial might cue to pitch. However we know that the pitch remains at 100 Hz regardless removal of 100 Hz partial. This is known as \textit{pitch of the missing fundamental} [5,6]

\subsubsection{Partials spacing model}
Now take the same sawtooth waveform, the first partial removed and consider its partials: 200 Hz, 300 Hz, 400 Hz, 500 Hz and so on. We know that the lowest frequency changed but the pitch remained at 100 Hz. What also remained as 100 Hz is the spacing between the frequencies of the partials. Hence one would assume that the average spacing between the partials might cue to pitch.

Now consider a square wave with the known partials of 100 Hz, 300 Hz, 500 Hz, 700 Hz and so on. The pitch of this waveform is known to be 100 Hz. But the spacing between the partials is 200 Hz. Hence the spacing model fails here.

\subsubsection{Greatest common divisor model}
Now consider the square wave again. The only connection between the pitch of the square wave and the frequency components is that the greatest common divisor of the frequency components is 100 Hz. This model predicts the pitch of the square wave, alongside fitting all the other waveforms proposed previously.

Now consider a complex, and inharmonic, waveform of following frequencies: 930 Hz, 1770 Hz, 2730 Hz, 3570 Hz, 4530 Hz. This is a slightly altered version of the harmonic complex with 900 Hz, 1800 Hz, 2700 Hz, 3600 Hz, 4500 Hz, each component is increased or decreased by 30 Hz. The harmonic (latter) tone clearly has a pitch corresponding to 900 Hz. Although the timbre changes, the inharmonic complex’s (former) pitch is also heard the same as 900 Hz. However, it clearly has e period of 30 Hz since this is the greatest common divisor of the given frequencies. Due to frequency uncertainty in hearing, together with the fact that partials’ effect on pitch diminish as harmonic number increases and the given partials need to be interpreted as $31^{th}$, $59^{th}$, $91^{th}$ (and so on) harmonics for a 30 Hz pitch candidate; the given partials are treated as $1^{th}$, $2^{nd}$, $3^{rd}$ (and so on) within-bandwidth-harmonics of 900 Hz. Hence the GCD model fails here. A pitch perception model should take the frequency uncertainty into account and interact with spectrograms with non-sharp peaks, rather than discretely defined frequency components.

\section{Spectral Models and MASP}
\subsection{Subharmonic Histogram}
Cheveigne A. [1] summarizes pattern-matching models of pitch perception as follows (see Fig. 1): “… This brings us to a final algorithm. Build a histogram in the following way: for each partial, find its subharmonics by dividing its frequency by successive small integers. For each subharmonic. Increment the corresponding histogram bin. Applied to the spectrum in Figure 1E, this produces the histogram bin. Applied to the spectrum in Figure 1E, this produces the histogram illustrated in Figure 1F. Among the bins, some are larger than the rest. The rightmost of the (infinite) set of largest bins is the cue to pitch. This algorithm works for all the spectra shown. It illustrates the principle of pattern-matching models of pitch perception.”
\begin{figure}[!t]
\centering
\includegraphics[width=3in]{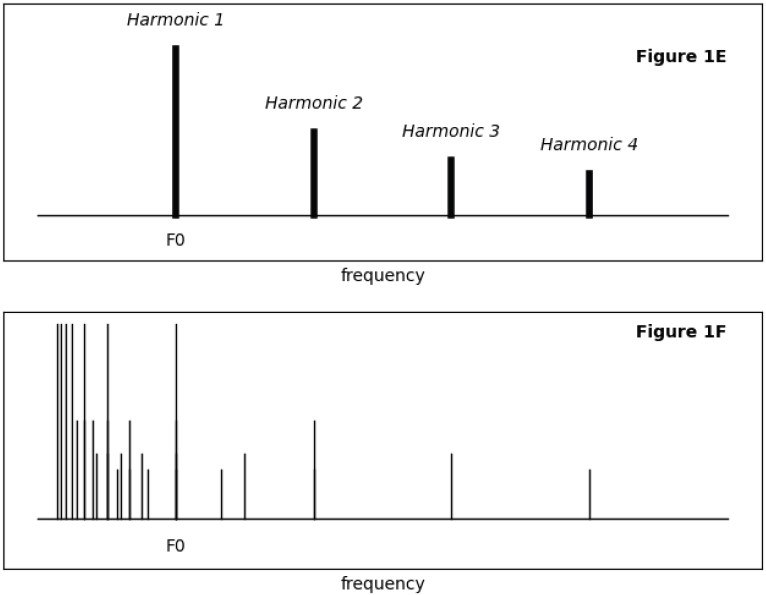}
\caption{A simple pattern-matching model described by Cheveigne [1].}
\label{fig_1}
\end{figure}

Although it seems to work for all the problematic cases in which the previously proposed pitch perception models suffered, this model lacks some properties for wide use:
\subsubsection{Too many peaks}
Although the pitch is explicitly defined as "rightmost of the highest peaks", it is neither a clear mathematical definition of the pitch nor has the rigor to explain why. The pitch of a good harmonic tone, which can be almost any note played on most of the musical instruments, is well defined and single. The output of the proposed model however is not as clear as the pitch is heard, for it has many pitch candidates and requires further human interpretation for the evaluation of the pitch.
\subsubsection{Discrete definition of frequency}
Although the visuals are for demonstration purposes, it can be risky to work with idealistic models as the bandwidth (i.e. uncertainty) of the frequency can be crucial to pitch perception: Altering the frequencies of the partials of a complex tone will not change the perceived pitch significantly since hearing is not that precise, but a discrete model like this will result in subharmonics not adding up for slightly altered harmonics.

\subsection{Flaw of Subharmonic Histogram}
As Helmholtz [7] explains, a string responds to all the frequencies that are integer-multiples of its fundamental. According to the pattern matching model proposed above, the output graph can be interpreted as the amount of vibrations of all the strings with fundamental resonance frequencies spanning the frequency range, in response to a complex tone.

This logic is reverse: with the strings of a fixed fundamental frequency that produce a perfect harmonic tone with partials that are integer-multiples of the fundamental, the spectrum is a result of the fundamental frequency of the string, not that the fundamental of the string is a result of the frequency spectrum. Hence the object must be to find the fundamental of an imaginary string by which the frequency spectrum might have been produced, rather than the strings that respond to that frequency spectrum.

The logical correspondence of this thought is: the pitch (fundamental frequency) must exist in all subharmonic series of all the partials of the complex. Hence the partials must be linked by \textit{and} rather than \textit{or}. The mathematical correspondence of this logic is that the subharmonic histograms must be multiplied rather than added.

\subsection{Harmonic Product Spectrum}
Thankfully this logic was applied by Schroeder [8] with Harmonic Product Spectrum, which multiplies the subharmonic spectra (1).
\begin{equation}
\label{deqn_ex1a}
Y(\omega) = \prod_{k=0}^{n} X(k\omega)
\end{equation}
where X($\omega$) is frequency spectrum, Y($\omega$) is pitch spectrum. As Camacho [9] points out, this formula fails whenever a harmonic is missing since it directly cancels out all the multiplication, by multiplication by zero. Another problem is that there is no logical reason of a strictly defined number of harmonics n, but defining it ‘up to infinity’ or any high number that allows encapsulation of all the hearing range is also going to lead infinitesimally small product results, not to mention that each partial contributes equally to the pitch, hence we also get peaks at subharmonics of the fundamentals as a result of equal contribution from partials.
\subsection{Mean Averaging Smoothed Product Spectrum}
Here we propose the MASP spectrum. 
\subsubsection{Smoothing the factors}
The underlying problem with the HPS is that, for it is a product, each subharmonic spectrum, which are the factors of the product, have too much effect on the product. Now consider frequency spectrum $X(\omega)$. For the multiplicative identity is a uniformity, let us define:
\begin{equation}
\label{deqn_ex1a}
a = \frac{1}{1+k^b} , \langle X \rangle = mean[X(\omega)]
\end{equation}
\begin{equation}
\label{deqn_ex1a}
X_k(\omega) = a  X(k\omega) + (1-a)  \langle X \rangle
\end{equation}
Where $k$ is harmonic number $X_k$ is the smoothed subharmonic spectrum of order $k$. Parameter $b$ may later be adjusted; $ b >> 1$ emphasizes lower harmonics and increases the possibility of encountering the first partial, whereas a lower $b$ (e.g. $b=0.25$) makes it possible to emphasize a low fundamental that does not exist in partials. By ensuring that $a$ starts from $0.5$ and approaches $0$ with bigger $k$, it is guaranteed that the reduced spectrum $X_k$ is never zero and asymptotically approaches to a uniform distribution at average value of $X(\omega)$, with increasing harmonic number.

This allows, provided normalization, for much higher iterations of $n$; for however higher the $n$, the more uniform $X_k$ becomes and the effect of multiplication becomes minimal. Hence the pitch spectrum implicitly (4) and explicitly (5) becomes;
\begin{equation}
\label{deqn_ex1a}
\bar{Y} = \prod_{k=0}^{n} X_k(\omega)
\end{equation}
\begin{equation}
\label{deqn_ex1a}
\bar{Y}(\omega) = \prod_{k=0}^{n}            [\frac{1}{a+k^b}X(k\omega) = \frac{k^b}{1+k^b} \langle X \rangle]
\end{equation}
\begin{figure}[!t]
\centering
\includegraphics[width=3in]{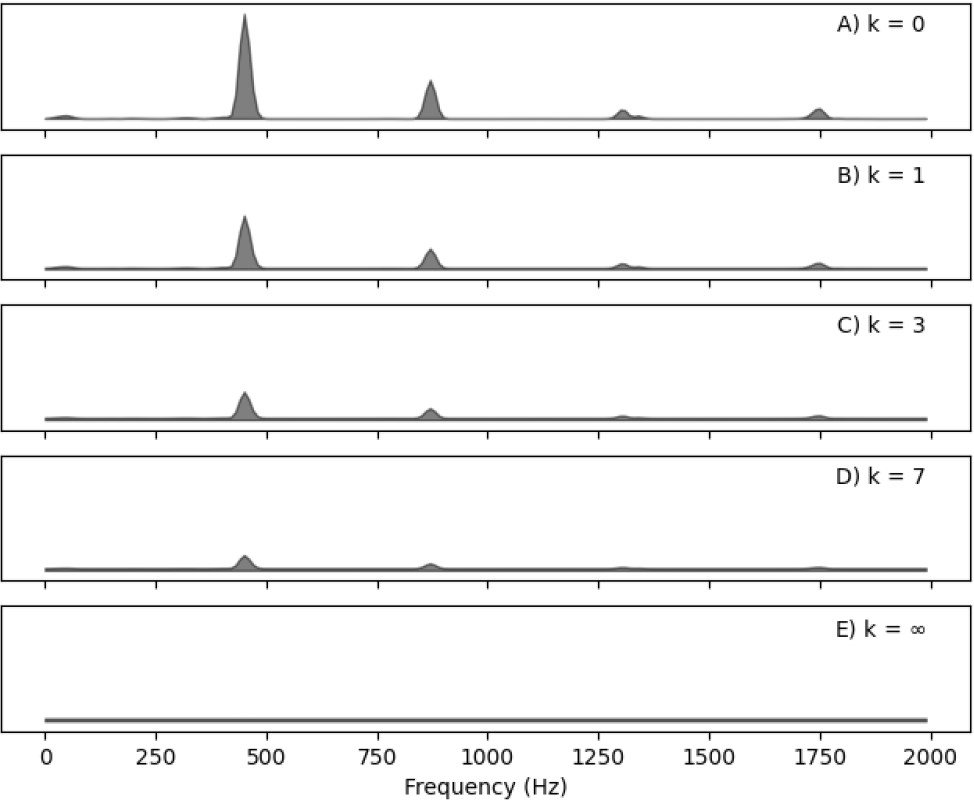}
\caption{The effect of mean-averaging smoothing on a spectrum, original spectrum on top. As $k$ increases, the spectrum approaches to become uniform.}
\label{fig_1}
\end{figure}
\subsubsection{Normalization by loudness}
Although the previous steps might allow us to keep the overall shape of the spectrum when encountered by zero products, and we get the desired peak at the desired frequency for the pitch spectrum; we still need to modify the amplitude to a plausible value. Here we are attempting to make this modification such that the pitch spectrum is also a representation of the loudness of the waveform:
\begin{equation}
\label{deqn_ex1a}
S = \frac{    log[1 + \sum X(\omega)]     }{      \sum \prod_{k=1}^{n} X_k (\omega)      }
\end{equation}
\begin{equation}
\label{deqn_ex1a}
\bar{Y} =S \prod_{k=0}^{n} X_k(\omega)
\end{equation}
Provided that $X(\omega)$ is a power spectrum, which can be obtained by absolute square of FT of the waveform.

\subsubsection{Logarithmic MASP spectrum}
A challenging part of pitch perception models is that spectra attained with a Fourier Transform is not suitable to model human perception. Frequency uncertainty in hearing is not constant but correlated with the frequency, i.e., bandwidths get wider as frequency increases and the bandwidth is only constant on a logarithmic scale. FT, on the contrary, creates spectra with a constant bandwidth throughout the frequency range. Oppenheim and Magnasco [10] also have shown that accuracy in human hearing can beat Fourier uncertainty.

A \textit{Constant Q Transform}, at the expense of a higher computational cost compared to FFT, creates much more suitable spectra to model perception since the bandwidths are constant on a logarithmic frequency scale. Hence it would be appropriate to formulate the MASP spectrum for logarithmic frequency scale for further use under CQT. Now consider a modified \textit{Mel scale} and a new frequency spectrum $F(m)$ defined on \textit{Mel scale}:
\begin{equation}
\label{deqn_ex1a}
m = log_2 \frac{\omega}{\omega_0} , F(m) = X(\omega_0 2^m) = X(\omega)
\end{equation}
where $\omega_0$ is an arbitrary reference frequency, ideally lowest on the desired scale: $m$ giving \textit{octave} distance with respect to the note associated to $\omega_0$. Multiplications on $\omega$ can be converted to \textit{Mel scale} as follows:
\begin{equation}
\label{deqn_ex1a}
\begin{split}
\frac{\omega_2}{\omega_1} = k \rightarrow m_2 - m_1 = log_2 k \\
X(k\omega) = F(m+log_2 k)
\end{split}
\end{equation}
Where a downsampling of $X$ by $k$ in linear scale became a shift of $F(m)$ by $log_2 k$ on \textit{Mel scale}. The smoothing defined in (3) should also be applied on $F$. Hence the desired $k^{th}$ subharmonic spectrum of $F(m)$ becomes:
\begin{equation}
\label{deqn_ex1a}
F_k(m) = a F(m + log_2k) + (1-a) \langle F \rangle
\end{equation}
where $\langle F \rangle$ is the mean of $F(m)$, $a$ and $(1-a)$ are weights defined in (2). Now consider a new pitch spectrum $P(m)$ defined on \textit{Mel scale}. The same principles between $X(\omega)$ and $Y(\omega)$ defined in (7) also applies between $P(m)$ and $Y(m)$. Hence $P(m)$ becomes;
\begin{equation}
\label{deqn_ex1a}
P(m) = \prod_{k=1}^{n} F_k(m)
\end{equation}
Here in (11), $F_k$ is the $k^{th}$ subharmonic spectrum of $F$, smoothed by an order of $k$, as shown in (10).

\subsection{Applications of MASP}
The MASP algorithm was implemented in Python using Librosa module for CQT computation. Full implementation may be seen on Reference [11].

METHOD: The algorithm was tested on a variety of synthetic and real musical instrument tones. The CQT was computed for the audio signal throughout the duration to create a CQT matrix of $density$ vs $frequency\times time$ to convert to another matrix of same dimensions that describes MASP spectrum. The MASP spectrum was derived using (11).

\subsubsection{Artificially generated tones}
For a sawtooth wave with its first and second harmonics removed and a aquare wave with its fundamental removed, MASP correctly identified the fundamental pitch (Fig. 3, 4).
\begin{figure}[!t]
\centering
\includegraphics[width=3in]{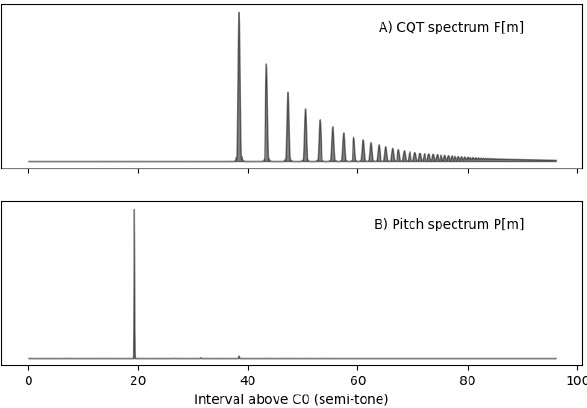}
\caption{Frequency (above) and MASP (below) spectra of a modified sawtooth wave with first and second partials removed}
\label{fig_1}
\end{figure}\begin{figure}[!t]
\centering
\includegraphics[width=3in]{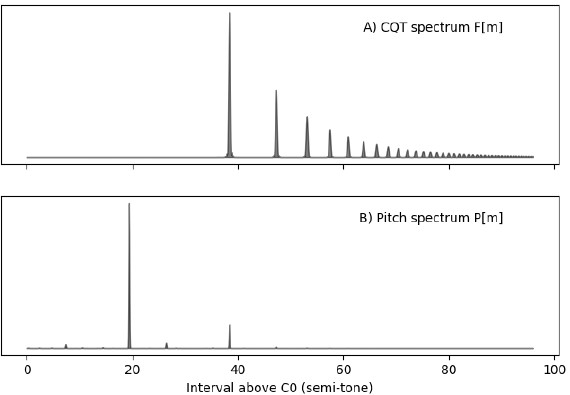}
\caption{Frequency (above) and MASP (below)  spectra of a modified square wave with first partial removed}
\label{fig_1}
\end{figure}
For an inharmonic complex with partials at 930, 1770, 2730, 3570, 4530, 5370 Hz -whose harmonic counterpart has partials at 900, 1800, 2700 Hz...-, MASP correctly indicated at a pitch of 900 Hz, avoiding the 30 Hz GCD pitfall (Fig. 5).
\begin{figure}[!t]
\centering
\includegraphics[width=3in]{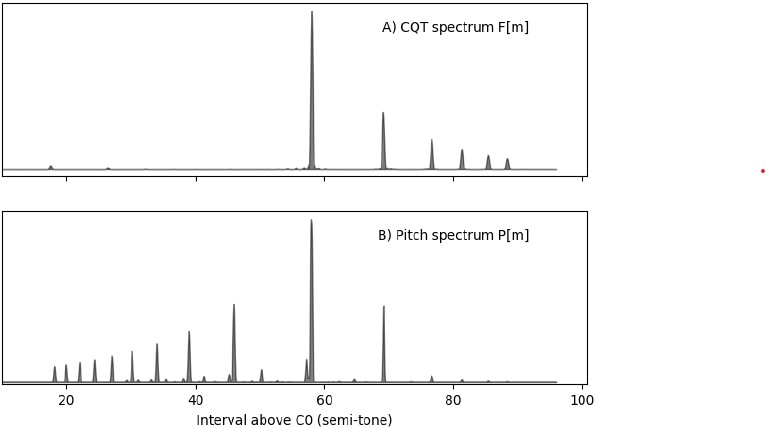}
\caption{Frequency (above) and MASP (below) spectra of a slightly inharmonic complex with partials 930, 1770, 2730... Hz}
\label{fig_1}
\end{figure}
For an inharmonic complex with partials at 900, 1100, 1300, 1500, 1700 Hz, MASP spectrum exhibited several pitch candidates, two highest of which are about 217 and 186 Hz, as claimed by Rasch \and Plomp for this ambiguous spectrum [4].
\begin{figure}[!t]
\centering
\includegraphics[width=3in]{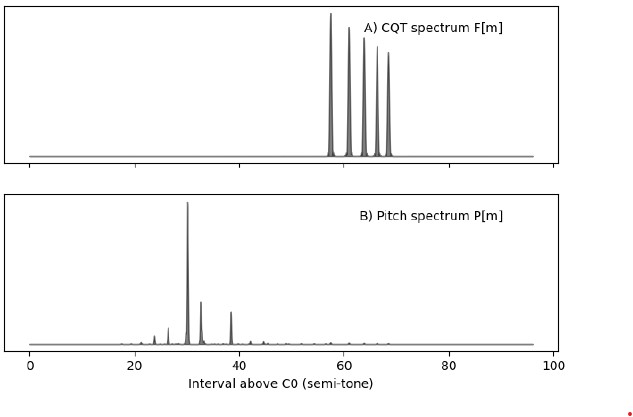}
\caption{Frequency (above) and MASP (below) spectra of a slightly inharmonic complex with partials 930, 1770, 2730... Hz}
\label{fig_1}
\end{figure}
\subsubsection{Real musical instruments}
MASP was applied to various musical tones such as a stable A4 = 440 Hz played on \textit{celli} (Fig. 7), a trumpet except (Fig. 8), and an oboe melody (Fig. 9), for most of which perceptually plausible pitch estimations attained.
\begin{figure}[!t]
\centering
\includegraphics[width=3in]{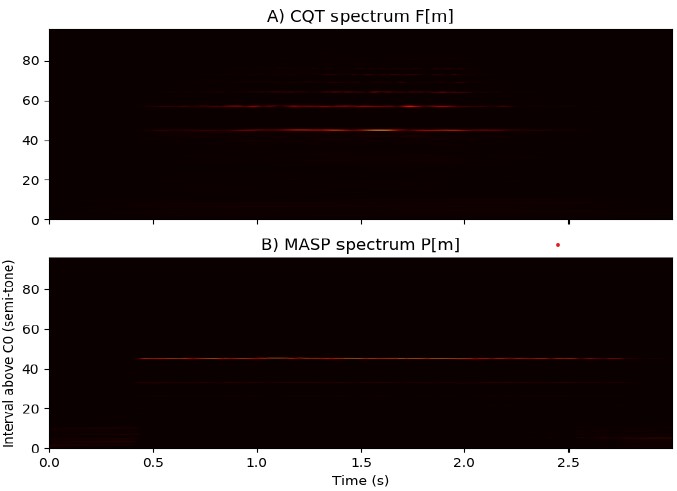}
\caption{Frequency (above) and MASP (below) spectravs time for a celli section playing A4 = 440 Hz}
\label{fig_1}
\end{figure}
\begin{figure}[!t]
\centering
\includegraphics[width=3in]{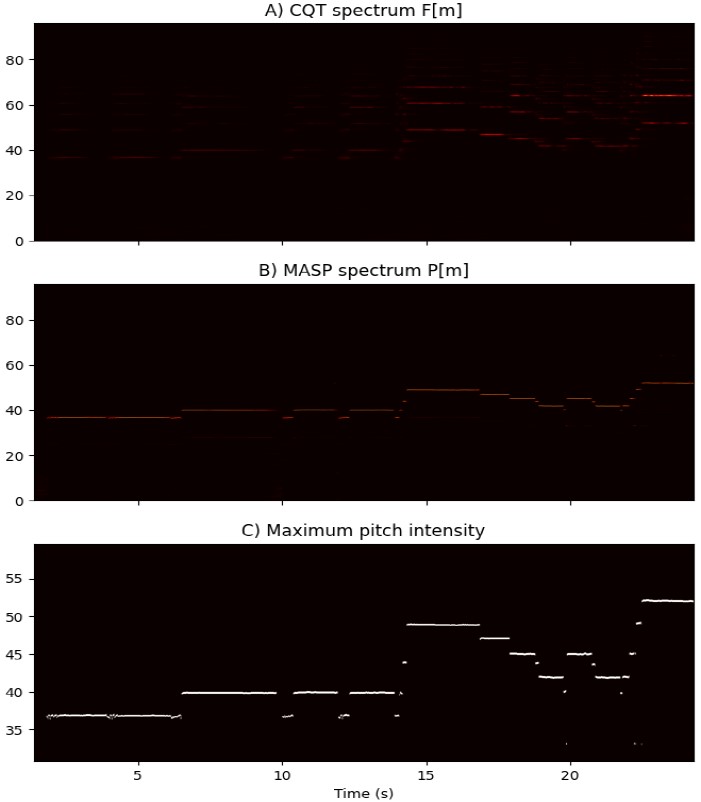}
\caption{Frequency (above) and MASP (center) spectra and maximum pitch density (below) vs time for a trumpet excerpt from Mahler's Symphony no. 5}
\label{fig_1}
\end{figure}
\begin{figure}[!t]
\centering
\includegraphics[width=3in]{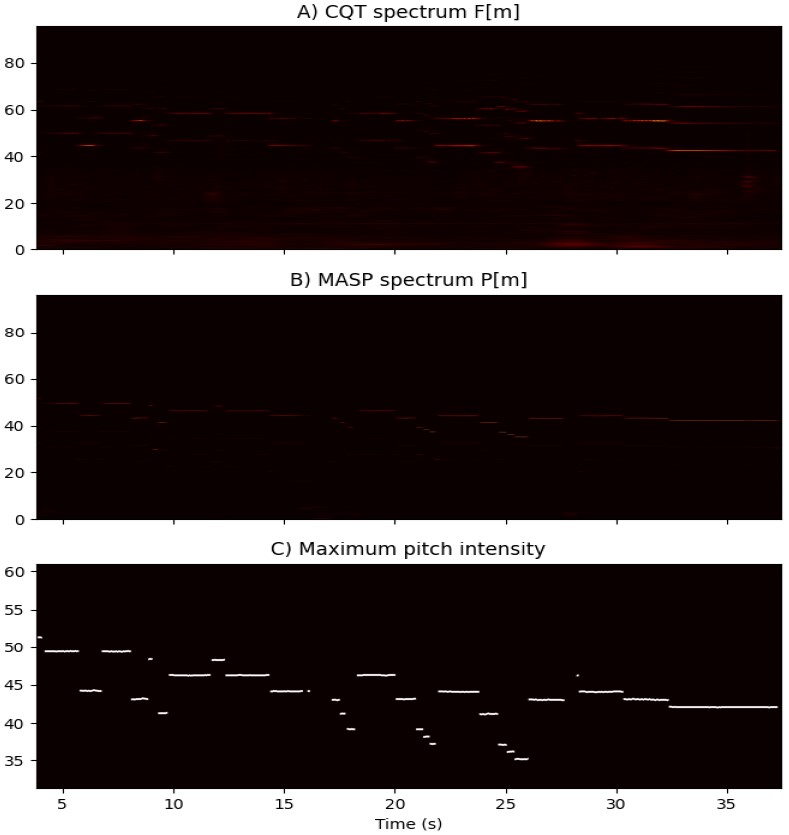}
\caption{Frequency (above) and MASP (center) spectra and maximum pitch density (below) vs time for an oboet melody from Wagner's Tristan und Isolde}
\label{fig_1}
\end{figure}
\section{Musical Consonance}
Musical consonance is a property, or quality, of two musical tones that describes their pleasantness or harmoniousness, when they are sounded together. It is an integral part of Western music, as making more than one separate voices sound harmonious is what puts Western music apart relative to its monophonic counterparts.

The most primitive theory of consonance was supposedly proposed by Pythagoras, which states that the simpler the ratio between frequencies of two tones that make up an interval, the more consonant the interval. Pythagoras used the consonant intervals to construct the primary musical scales. For monophonic tones are also played on scales, one would assume that consonance is actually not important only for homophony, but also a part of the foundation of monophonic music.
\subsection{Consonance Theories}
Apart from the primitive Pythagorean version of 'simple ratios', many theories have been proposed to estimate consonance and rank musical intervals with respect to their consonances, all of which in some way or other are related to simple-integer-ratios.
\subsubsection{Discrete theories}
Euler proposed \textit{gradus suavitatis} or "degree of softness" [12] where the lower the \textit{Least Common Multiple} of $a$ and $b$ ($\frac{a}{b}$ is the integer frequency ratio of the tones in question in simplest form), the more harmonious the interval is. Likes of Stolzenburg [13] and Brefeld [14] followed Euler's approach and proposed more sophisticated measures of consonance (See Stolzenburg, 2015, pg. 4)
\subsubsection{Spectral theories}
Discrete theories give reasonable, and experimentally consistent, results for consonance evaluation for integer-ratio-intervals. When it comes to evaluation of inharmonic intervals (like the \textit{perfect fifth} in 12-tone equal temperament, whose ratio is $2^{7/12}$ which is a tiny bit smaller than $\frac{3}{2}$) most of the discrete theories round the slightly inharmonic intervals, giving no quantitative basis of how much rounding is allowed and what effect slight deviations from simple-integer ratios exactly have.

A prominent spectral theory of consonance came from Plomp and Levelt [15] where dissonance (commonly used as inverse of consonance) is related to roughness, which occur when two pure tones are too close relative to critical bandwidth. Applying the principle to two harmonic complex tones with many partials, it is shown that the overall roughness, which is attained by summing all roughness between all binary combination of two tones' partials, is a good measure of dissonance between two tones.

A pitfall of Plomp and Levelt's roughness approach is pure tones. As it depends upon the existence of higher harmonics to assess overall roughness, the roughness curve of two sinusoids with respect to relative frequency deviation vanishes at zero and infinity, and peaks at an intermediary point, depending on critical bandwidth, which is uninteresting in terms of a musical interval. This would suggest that a concept of musical consonance for pure tones does not exist and a consonance estimation only becomes meaningful when multiple harmonics are involved.
\begin{figure}[!t]
\centering
\includegraphics[width=3in]{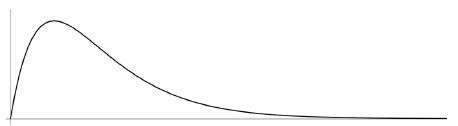}
\caption{A roughness curve depending on frequency separation. Note that most roughness curves are roughly the same (similar) and scaled depending on critical bandwidth}
\label{fig_1}
\end{figure}
\subsubsection{Periodicity}
Generally, consonant intervals of Western music exhibit high periodicity [16]. Hence many theories have been proposed to rely consonance upon periodicity of the joint waveform that two tones composes, most of which also give accurate-with-experience results. A version of Euler's consonance evaluation by $LCM(a,b)$ may be made up: instead of $LCM$, compute $GCD$ of the frequencies of the tones. As discussed in \textit{Section II.B.III}, $GCD$ of two tones gives a strong cue to the periodicity of the joint waveform.

Although the procedure is much more complex than a periodicity measure, Trulla \textit{et al} [17] proposed a consonance assessment that gives a reasonable consonance curve for pure tones, which Plomp and Levelt's model lacks.
\subsection{Consonance Evaluation with MASP Spectrum}
Motivated by the correlation between periodicity and consonance shown in previous works in the literature, and given another even stronger correlation between periodicity and pitch: one of the first things one should do for a pitch detection algorithm is to use it to evaluate musical consonance.
\subsubsection{MASP spectrum of intervals with pure tones}
Now consider a waveform of two pure tones that lasts 14 seconds starting from $-1^{th}$ second. One of the tones is kept at a constant frequency and the other is logarithmically incremented, meeting the frequency of the first tone at $t = 0$ and reaching twice that frequency at $t = 12 s$, so that a transition of an \textit{octave} corresponds to 12 seconds and each second corresponds to a \textit{semitone} increment of the second tone.
\begin{figure}[!t]
\centering
\includegraphics[width=3in]{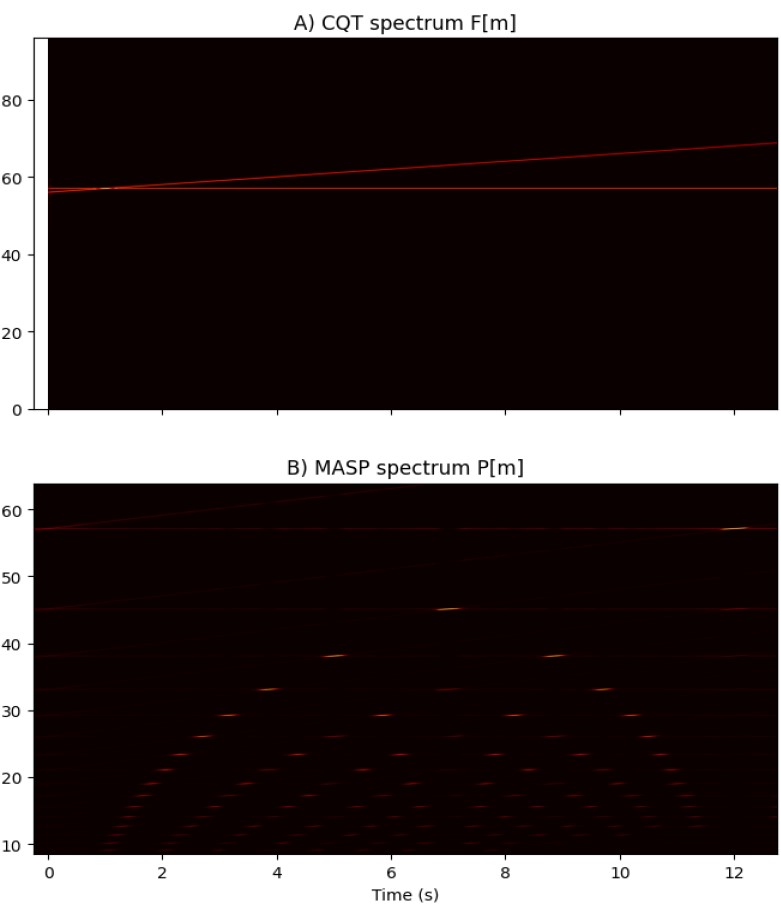}
\caption{Frequency \& MASP spectra vs time for two pure tones. First tone held constant, second tone increments a semitone per second to cover (-1,12) semitone range relative to the first, spanning an octave.}
\label{fig_1}
\end{figure}
Looking at Fig. 11 from top to bottom: The first bright area occurs around $t=12 s$, corresponding to 12 \textit{semitones} or \textit{octave}. The second bright area occurs around $t=7$, corresponding to 7 \textit{semitones} or \textit{perfect fifth}. The third(s) bright areas occurs around $t=5$ and $t=9$, corresponding to 5 and 9 \textit{semitones} or \textit{perfect fourth} and \textit{major seventh}.
The list may continue and the consonant intervals may be seen as bright regions on MASP spectrum, with arguably the more consonants having higher locations. Those patterns occur with complex tones too, but much weaker. While the parameters were $b=0.5$ and $factorscale=0.5$ (an argument of Librosa's CQT function that defines time width, which is inversely proportional to frequency bandwidth due to \textit{uncertainty principle}) for the pure wave example, we had to change both to $0.1$ to see a similar pattern with a sawtooth wave (Fig. 12).
\begin{figure}[!t]
\centering
\includegraphics[width=3in]{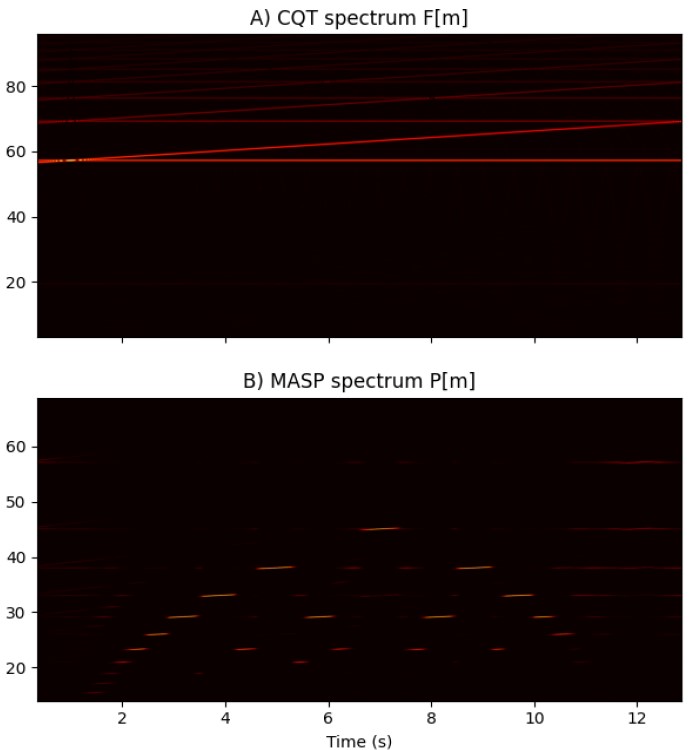}
\caption{Spectra with the same tones as in Fig. 11, but with sawtooth waves instead of pure tones}
\label{fig_1}
\end{figure}
An important property about this supposed consonance plot is that the bright regions do not suddenly vanish after passing the integer-ratio points. For instance; for a point that corresponds to a ratio of $1.49$, the point is still almost as bright as $1.5$, giving higher room of errors for frequency fluctuations. This is important regarding Western prominent 12-tone equal temperament, since none of the intervals, but octave, are not integer ratios but rather irrational.
\subsubsection{Expectancy of MASP spectrum}
Fig. 15 \& 16 give an insight, that is close to consonance. But in order to rank intervals with respect to how consonant they are, one should formulate a scalar quantity.
The frequency spectrum and MASP spectrum $F(m)$ and $P(m)$ at a given time, are a functions of frequency (on \textit{Mel scale}), $m$. Let us assume that $F(m)$ and $P(m)$ are strictly normalised (\textit{strictly} here means that they are scaled to make the sum 1, to not to be confused with the normalisation defined on \textit{Section III.D.II}) so that probabilistic operations can be done on them. Let $\mathbf{m}_F$ and $\mathbf{m}_P$ be the random variables associated with probability distributions $F(m)$ and $P(m)$. Since more consonant intervals seem to be exhibiting greater peaks at higher $m$ values, also we would like to focus on relative positions of the bright parts with respect to the frequency spectrum; we are tempted to define a harmonicity measure $H$ as follows:
\begin{equation}
\label{deqn_ex1a}
log_2H = \langle \mathbf{m}_F \rangle - \langle \mathbf{m}_P \rangle
\end{equation}
where $\langle \mathbf{m}_F \rangle$ and $\langle \mathbf{m}_P \rangle$ are \textit{expected values} of the variables for their corresponding distributions $F(m)$ and $P(m)$.

An $H$ curve for the same waveform (in Fig. 11) exhibits suspected peaks in consonant intervals (Fig. 13). The effect of roughness may be seen as wild fluctuations around the \textit{unison} point, $t=0$ in the $H$ curve. As the tones are pure, this effect is only seen around \textit{unison}, for there are no partials to coincide elsewhere. However this is not the case for a complex waveform as fluctuations happen almost everywhere (Fig. 14).
\begin{figure}[!t]
\centering
\includegraphics[width=3in]{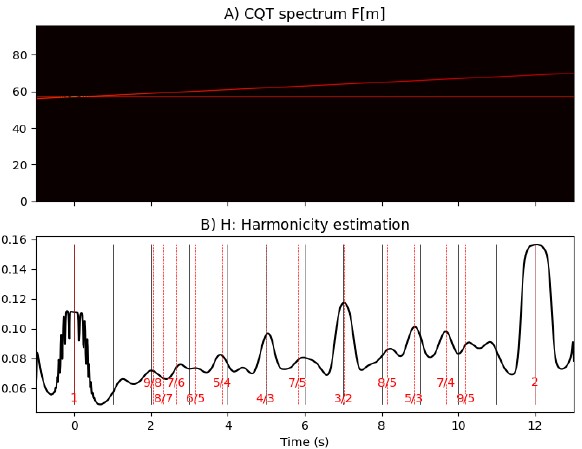}
\caption{Freq. spectrum \& $H$-value vs time (semitone) for tones in Fig. 11}
\label{fig_1}
\end{figure}
\begin{figure}[!t]
\centering
\includegraphics[width=3in]{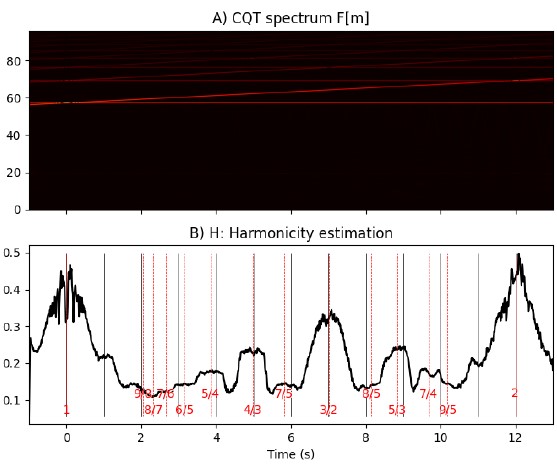}
\caption{Freq. spectrum \& $H$-value vs time (semitone) for tones in Fig. 13}
\label{fig_1}
\end{figure}
Although the estimated $H$-values differed between pure and complex tone cases, where $H$-values for complex tones turned out to be higher; relative $H$-values for intervals stayed similar in both cases. The \textit{unison}, which is by definition the most consonant interval, performed lower than \textit{octave} and \textit{perfect fifth}. The reason is as follows: When there are more than one partials in the frequency spectrum and if the partials are harmonically related, they overlap in the product of subharmonic spectra to exhibit a sharper peak in the MASP spectrum. When there is one partial however, the subharmonics never overlap and make sharper peaks to yield a strong pitch. As a waveform with one tone is practically the same with one with two tones that yield the same pitch, consonance estimation at \textit{unison} for multiple tones become irrelevant with this approach where it is inevitable to treat two separate tones as one at \textit{unison}.

To estimate consonance between two tones, one may simply separate the tones and constitute MASP spectrum. Suppose that $X_1(\omega)$ and $X_2(\omega)$ are linear-domain (not Mel scale) frequency spectra of two distinct tones that are sounded together, $\bar{Y}_1$ and $\bar{Y}_2$ are MASP spectra of those distinct tones, respectively. The strictly normalised joint MASP spectrum may be obtained by;
\begin{equation}
\label{deqn_ex1a}
\bar{y}(\omega)=\bar{Y}_1 (\omega) \bar{Y}_2 (\omega) , \bar{Y}(\omega)=\frac{    \bar{y}(\omega)    }{    \sum \bar{y}(\omega)   }
\end{equation}
and the normalized joint frequency spectrum by;
\begin{equation}
\label{deqn_ex1a}
x(\omega)=X_1 (\omega)+X_2 (\omega) , X(\omega)=\frac{    x(\omega)    }{    \sum x(\omega)   }
\end{equation}
Now assume that $\mathbf{\omega}_{\bar{Y}}$ and $\mathbf{\omega}_{X}$ are random variables associated with distributions $\bar{Y}(\omega)$ and $(\omega)$, respectively. Following the conversions in (9) and harmonicity measure defined in (12);
\begin{equation}
\label{deqn_ex1a}
H=\frac{ \langle   \mathbf{\omega}_{\bar{Y}}     \rangle}{ \langle    \mathbf{\omega}_{X}  \rangle}
\end{equation}
where $\langle   \mathbf{\omega}_{\bar{Y}}     \rangle$ and $\langle    \mathbf{\omega}_{X}  \rangle$ are \textit{expected values} of the variables for their corresponding distributions  $\bar{Y}(\omega)$ and  $X(\omega)$:
\begin{equation}
\label{deqn_ex1a}
\langle \mathbf{\omega}_{\bar{Y}} \rangle = \sum \omega \bar{Y}(\omega), \langle \mathbf{\omega}_X\rangle = \sum \omega X(\omega)
\end{equation}
For an $H$ curve of specified parameters, maxima are at simple-integer-ratios (Fig. 15).
\begin{figure*}[!t]
\centering
\includegraphics[width=6in]{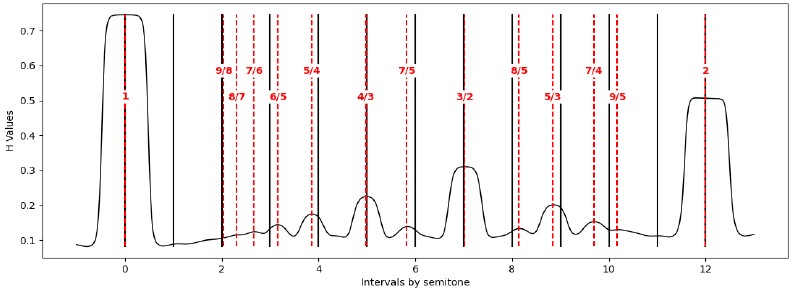}
\caption{$H$ curve for intervals within an \textit{octave}, following (15) with $b=0.5$. For frequency spectra, Gaussian distributions with $\sigma = 0.1\cdot semitone$ is used. Black verticals represent 12-tone equal temperament intervals, red verticals represent intervals with simple-integer-ratio.}
\label{fig_1}
\end{figure*}
\begin{table}[!t]
\caption{H-values for common musical intervals}
\label{tab:table1}
\centering
\begin{tabular}{|l|c|c|c|l|}
\hline
\textbf{12-tet Intervals} & \textbf{Step} & \textbf{\textit{H}-value} & \textbf{Ratio} & \textbf{Rational Intervals} \\
\hline
\textit{Unison} & 0 & 0.745 & 1 & \textit{Unison} \\
\hline
\textit{Octave} & 12 & 0.506 & 2 & \textit{Octave} \\
\hline
-- & -- & 0.310 & 3/2 & \textit{Perfect Fifth} \\
\hline
\textit{Perfect Fifth} & 7 & 0.310 &--  &-- \\
\hline
 --& --& 0.225 & 4/3 & \textit{Perfect Fourth} \\
\hline
\textit{Perfect Fourth} & 5 & 0.225 &--  &-- \\
\hline
  --&-- & 0.200 & 5/3 & \textit{Major Sixth} \\
\hline
\textit{Major Sixth} & 9 & 0.194 & --& -- \\
\hline
--&-- & 0.175 & 5/4 & \textit{Major Third} \\
\hline
\textit{Major Third} & 4 & 0.166 &--  &-- \\
\hline
-- & -- & 0.153 & 7/4 & \textit{Septimal Min. Seventh} \\
\hline
-- & -- & 0.144 & 6/5 & \textit{Minor Third} \\
\hline
-- &--  & 0.139 & 7/5 & \textit{Diminished Fifth} \\
\hline
 --& -- & 0.133 & 8/5 & \textit{Diminished Sixth} \\
\hline
\textit{Minor Third} & 3 & 0.133 & -- &--  \\
\hline
\textit{Minor Seventh} & 10 & 0.130 & -- &-- \\
\hline
\textit{Tritone} & 6 & 0.129 & -- &-- \\
\hline
\textit{Minor Sixth} & 8 & 0.126 & -- & -- \\
\hline
-- &--  & 0.124 & 7/6 & \textit{Septimal Minor Third} \\
\hline
\textit{Major Seventh} & 11 & 0.112 & -- & -- \\
\hline
\textit{Major Second} & 2 & 0.105 & -- & -- \\
\hline
\textit{Minor Second} & 1 & 0.089 & -- & --\\
\hline
\end{tabular}
\end{table}

It can be proven by (2), (3), and (4) that $H$-value for \textit{unison} approaches $1$ for the limit case as $b \to \infty$. But in this case, no subharmonics are involved and MASP spectra and frequency spectra become nearly identical, hence relevant for neither pitch nor consonance estimation.
Following a similar method used for two tones, H-values for three tones may also be attained for chords (Fig. 16)
\begin{figure}[!t]
\centering
\includegraphics[width=3in]{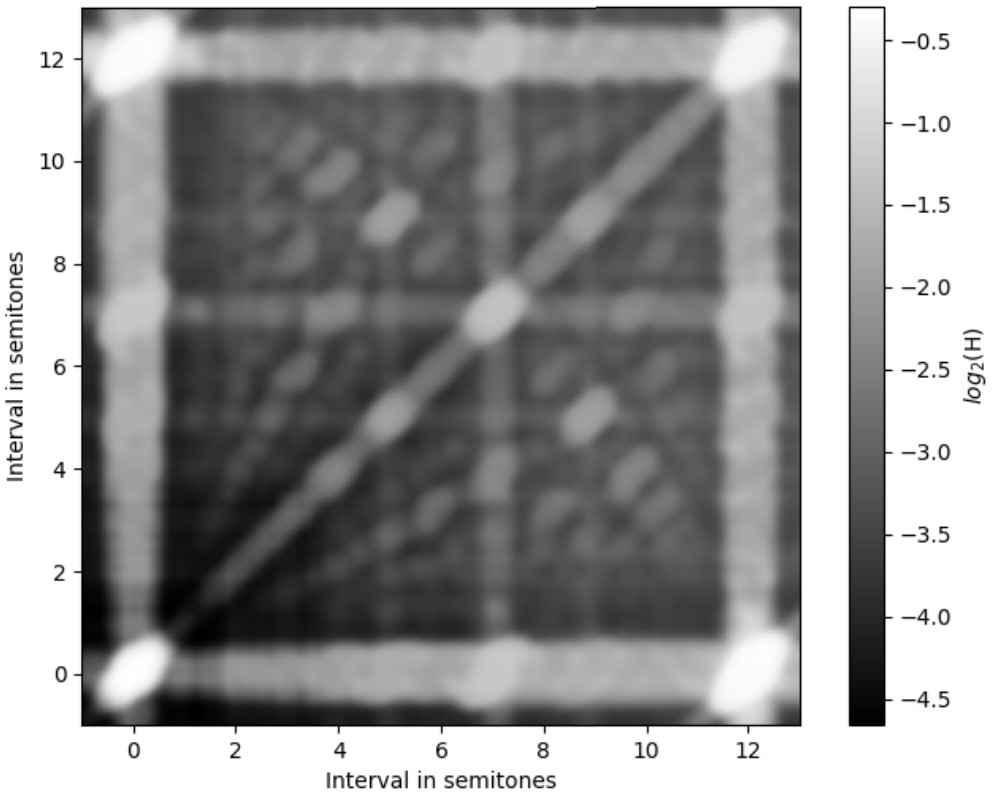}
\caption{$H$-values for three sinusoidal tones. One note held at 440 Hz, second \& third tones are varied along $x$ and $y$ axes, corresponding to specified semitone distances relative to the first tone.}
\label{fig_1}
\end{figure}

\begin{table}[!t]
\caption{H-values for common musical chords in 12-tone equal temperament\label{tab:table1}}
\centering
\begin{tabular}{|c||c||c|}
\hline
Sample notes & Interval in semitone & $H$-value \\
\hline
C, F, A & (0,5,9) & 0.257\\
\hline
C, E, G & (0,4,7) & 0.194\\
\hline
C, E, Bb & (0,4,10) & 0.182\\
\hline
C, Eb, Ab & (0,3,8) & 0.145\\
\hline
C, Eb, G & (0,3,7) & 0.141\\
\hline
C, Eb, A & (0,3,9) & 0.119\\
\hline
C, E, A & (0,4,9) & 0.115\\
\hline
C, Eb, Gb & (0,3,6) & 0.110\\
\hline
\end{tabular}
\end{table}
In Fig. 16, the bright outer square and the bright $x=y$ diagonal are uninteresting regions since two notes are same there. In the remaining region the brightest spot ($x\sim 9, y\sim 5$ and $x\sim 5, y\sim 9$) corresponds to \textit{second inversion} of \textit{major triad}. The $H$-values for remaining known chords may be seen in \textit{Table II}.
\section{Discussion \& Limitations}
We would like to treat \textit{consonance} and \textit{dissonance} as separate qualities, instead of just being opposite of one another. As annoying as a mis-tuned \textit{unison} with beating can be, it still yields a clear pitch hence consonant just with a different \textit{timbre}. What separates the timbre of a \textit{solo violin} and a \textit{violin section} is that timbre, caused by many frequency errors done by many violinists of a section, yet the sound still yields a perfect pitch perception and harmonicity. Furthermore, roughness-based theories of consonance rely upon existence of higher harmonics and no sense of consonance can be established for two pure tones. Hence, we would like to call roughness-based qualities of tones as "dissonance" and periodicity based qualities as harmonicity or "consonance".

As is the case with roughness based approaches, there is no one true $H$ curve as it depends upon frequency bandwidth and different $b$ values -see (2)-. These, for a listener, might depend upon music education and cultural exposition. If perception of consonance is truly linked to periodicity or pitch perception, one's perception of pitch might also slightly vary depending upon whether one is accustomed to monophonic or homophonic music.

The algorithm has not been applied to polyphonic cases as it is only suitable for monophonic cases. We also believe that such an algorithm that is applicable for multiple pitch detection should be both spectral and temporal, i.e., a pitch evaluation at a point of time of a waveform should be affected by spectra of the preceding points of time.

Let us explain why: suppose \textit{Violin I} is playing a note, \textit{Violin II} is also playing, at the same time, a note one octave above (two times the frequency) the \textit{Violin I}'s. For the spectrum of the joint waveform, there is no logical reason to state that two separate tones exist there since each harmonic of the second tone exists in the harmonics generated by the \textit{Violin I}’s fundamental. The same logic applies whenever the fundamentals of the two tones are linked with a simple-integer-ratio, as that condition will fit them into harmonic series of a lower fundamental: Suppose \textit{Violin I} is playing $A4 = 440 Hz$ and \textit{Violin II} is playing one \textit{perfect fifth} above ($E5 = 660 Hz$). This time all of the harmonics of both tones fit into the harmonic series of a lower fundamental of $A3 = 220 Hz$, hence any pitch perception algorithm that is claimed to be performing good under missing fundamentals should detect $220 Hz$ as a pitch candidate. It is only when they continue to play and make different note transitions that it becomes clear that there are actually two tones. Following the same logic, any complex harmonic actually is several sinusoids, the complex is always heard as one single tone when it is used for a musical melody, if all of the sinusoids change frequencies by the same fraction. Therefore most of the pitch perception algorithms that are designed to be applied at a fixed time interval are intrinsically based on the assumption that a future spectrum of the waveform, i.e., the next notes of a melody, are going to contain such frequencies that all of them will have changed by a certain ratio that corresponds to the note transition. A multiple pitch detector, hence, should analyze the \textit{cross-correlations} between the spectra at different time intervals of the waveform.

There is an exception to this, i.e., a true multiple pitch detector that is purely spectral. If the algorithm has \textit{a priori} information of \textit{inter se} ratios of amplitudes of harmonics in the separate tones, i.e., the \textit{timbres} of the tones; it could successfully detect multiple pitches. However this algorithm would not be a generalized one and would need to be updated whenever there is a new \textit{timbre} introduced. Also for a listener, that \textit{a priori} knowledge of \textit{timbre} is not truly \textit{a priori}, for one must be exposed to that tone previously in the lifespan, hence still a temporal part of pitch detection is involved.

\end{document}